\documentclass[onecolumn,aps,showpacs]{revtex4}

\usepackage{amsmath,amssymb,graphicx}

\begin{document}
\title{Random matrix ensembles: Wang-Landau algorithm for spectral densities}

\author{Santosh Kumar}
\email{skumar.physics@gmail.com}
\affiliation{                    
  Fakult\"at f\"ur Physik, Universit\"at Duisburg-Essen, Lotharstra\ss{}e 1, 47048 Duisburg, Germany
}

\begin{abstract}
We propose a method based on the Wang-Landau algorithm to numerically generate the spectral densities of random matrix ensembles.
 The method employs Dyson's log-gas formalism for random matrix eigenvalues and also enables one to simultaneously investigate the
 thermodynamic properties. This approach is a powerful alternative to the conventionally used Monte Carlo simulations based on the
 Boltzmann sampling, and is ideally suited for investigating $\beta$-ensembles.
\end{abstract}

\maketitle

\section{Introduction}

Random matrix theory (RMT) was introduced in physics by Wigner to describe the spectral properties of heavy nuclei. It is now a huge subject with applications in areas as varied as mesoscopic physics, quantum field theory, number theory, wireless communications and econophysics~\cite{Mehta2004,B1981,Forrester2010,GGW1998,Handbook2011}.

The statistics of eigenvalues plays a crucial role in the study of random matrix ensembles (RME). It not only encodes valuable information about a given ensemble but also enables one to calculate important observables related to the eigenvalues. Quite often one needs to evaluate the eigenvalue densities numerically. This is because the analytical expressions for the densities are available only for certain special ensembles and/or in asymptotic limits. In addition, the numerical simulation of spectra facilitates the verification of known analytical results. The conventional numerical schemes for generating eignvalue densities rely on either the diagonalisation approach using the matrix model whenever available, or implementing the Boltzmann-sampling-based Monte Carlo simulation~\cite{GPPS2003}. The latter uses Dyson's log-gas picture for the eigenvalues of RME~\cite{Mehta2004,Dyson1962}. We propose here a method based on Wang-Landau algorithm (WLA)~\cite{WL} which serves as a powerful alternative to these approaches. 

Our proposed method relies on calculating the microcanonical density of states of the Dyson log gas associated with the given random matrix ensemble. The density of eigenvalues or distribution of any related observable (function of eigenvalues) can be obtained from this information. RME are characterised by the Dyson index $\beta$ which plays the role of inverse temperature in the log-gas thermodynamics. The superiority of WLA over the conventional schemes lies in that one can evaluate the spectral density as well as the distributions of related observables for multiple $\beta$ values in a single simulation. The method, therefore, can be viewed as a natural tool in exploring the $\beta$-ensembles~\cite{DE2002,ABG2012,ABMV2012}. Moreover, WLA provides the additional advantage of producing density of states of the log gas during the simulation, which can be used to investigate its thermodynamic properties~\cite{Mehta2004,Dyson1962}.

\section{Wang-Landau algorithm}
\label{SecWLA}

Wang-Landau algorithm~\cite{WL} is a non-Boltzmann sampling method which was originally applied to standard statistical systems like Ising and Potts models which exhibit discrete energy states. WLA aims at estimating accurately the microcanonical density of states $g(E)$ (up to a multiplicative constant) by performing a random walk in the energy space. The key idea behind this algorithm is that if during the random walk the configurations $\chi$ with associated energy $E$ are sampled with a probability proportional to $1/g(E)$, the corresponding histogram $H(E)$ of visited configurations becomes flat. The density of states being unknown at the beginning of the simulation, one starts with an initial guess for $g(E)$ which is converged towards the true density in an iterative manner. Since the method gives direct access to the density of states of the system, which is independent of the temperature, one can calculate various thermodynamic averages by canonical reweighting at arbitrary nonzero temperature~\cite{WL}. There has been several improvements and refinements to the WLA by various authors~\cite{SDP2002,TD2005,ZB2005,GHH2007,BP2007,BMP2008}, consequently it has been successfully applied to systems possessing continuous energy spectra also, e.g., liquid crystals, polymers, biomolecules etc. See for example~\cite{GHH2007}, where the authors propose a hybrid algorithm based on the Wang-Landau and transition matrix Monte Carlo methods. In our calculations we implement the  $t^{-1}$ variant~\cite{BP2007,BMP2008} of WLA which facilitates, in general, a faster convergence rate than the conventional scheme.

From the information of $g(E)$ we can immediately evaluate the canonical probability at arbitrary nonzero temperature $T=1/(k_B\beta)$ as
\begin{equation}
P_{\beta,E}(E)=\frac{g(E)e^{-\beta E}}{\sum_{E'} g(E')e^{-\beta E'}},
\end{equation}
$k_B$ being the Boltzmann constant. The sum in the above equation runs over the sampled energy points in the energy window of interest. The canonical average of an observable depending explicitly on energy, say $Q(E)$, can then be easily calculated as
\begin{equation}
\langle Q \rangle=\sum_E Q(E) P_{\beta,E}(E).
\end{equation}
For example, the moments of energy $\langle E^n\rangle$ can be evaluated as $\sum_E E^n P_{\beta,E}(E)$, where $n=0,1,2,...$. By definition we have $\langle E^0 \rangle=1$. The average energy is given by the first moment, $U=\langle E^1\rangle$. The specific heat, which is related to the second cumulant of energy, is obtained using  $C_v=k_B\beta^2 \left(\langle E^2\rangle - \langle E^1\rangle^2\right)$.

For evaluation of the quantities which do not depend explicitly on $E$, say $A$, there are two ways one can proceed. The first approach is based on an extension, where instead of performing the random walk in only $E$-space, one performs the random walk in two-dimensional $(E,A)$ space with probability proportional to $1/\mathcal{G}(E,A)$, $\mathcal{G}(E,A)$ being the two-dimensional density of states~\cite{ZSTL2006}. The distribution of observable $A$ can be calculated at the end of simulation using
\begin{equation}
\label{PA}
P_{\beta,A}(A)=\frac{\sum_E \mathcal{G}(E,A)e^{-\beta E}}{\sum_{E} \sum_{A'} \mathcal{G}(E,A')e^{-\beta E}}~,
\end{equation}
which can be further used to calculate the average $\langle A \rangle$ using $\sum_A A P_{\beta,E}(A)$. However, performing the random walk in two-dimensional space is usually computationally expensive, being plagued with convergence issues etc.~\cite{ZSTL2006}. The alternative approach relies on estimating $g(E)$ first and then performing a production run~\cite{GWLX2009,WuL2012}. The production run comprises a post-simulation employing ``Wang-Landau resampling"  using the already converged $g(E)$. During this run, a histogram $\mathcal{H}(E,A)$ of the visits in the $(E,A)$ space is generated. The joint density $\mathcal{G}(E,A)$ can then be obtained using~\cite{GWLX2009}
\begin{equation}
\label{GEA}
\mathcal{G}(E,A)\sim g(E) \mathcal{H}(E,A). 
\end{equation} 
Eq. (\ref{PA}) can finally be used to study the desired observable.

\begin{figure*}[t]
\centerline{\includegraphics[width=.95\textwidth]{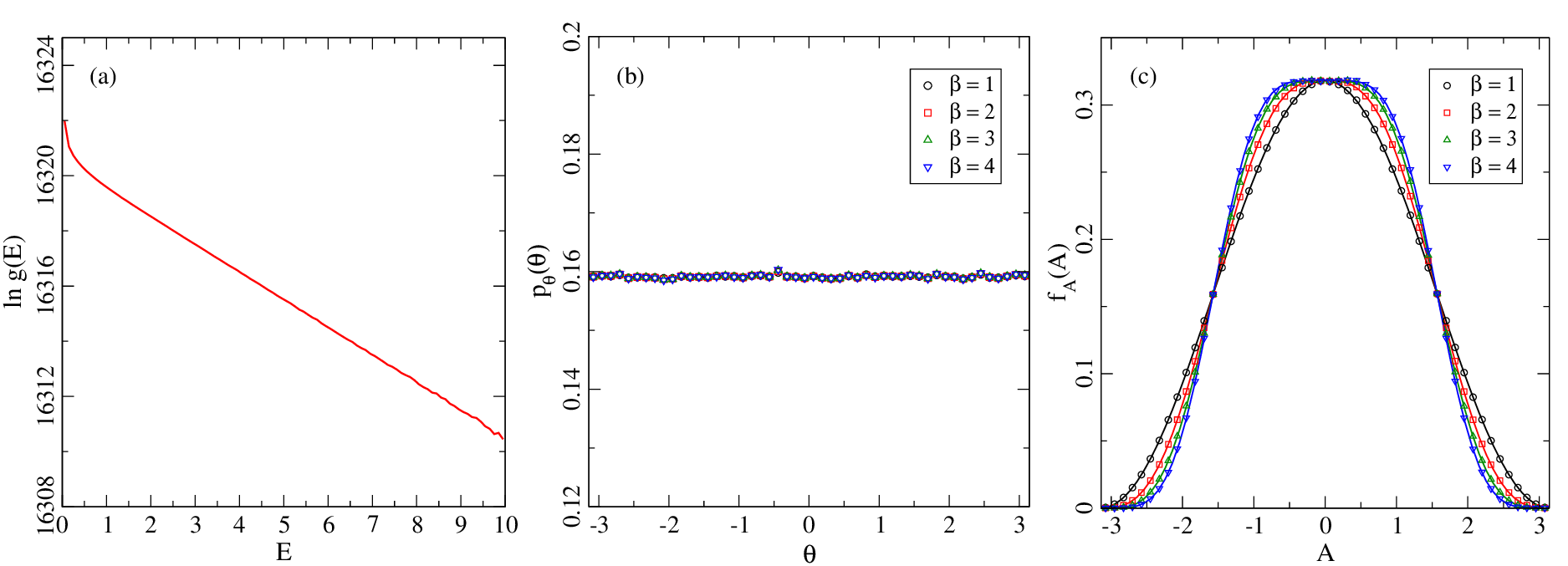}}
\caption[]{ (a) Log of the unnormalised density of states (b) Density of eigenphases for Dyson ensemble, eq.~\eqref{jpd_chi} with $w(e^{i\theta})=1$ (c) Density of observable $A=\sum_j\theta_j/N$. Uniform distribution $p_\theta(\theta)=(2\pi)^{-1}$ is obtained in (b) for each $\beta$ value. In (c) the symbols are the simulation results while the solid lines represent the analytical results given by eq.~\eqref{fA}.}
\label{Fig1}
\end{figure*}

\section{Dyson log gas}
\label{SecDLG}

Dyson in his pioneering work provided a physical picture to the eigenvalues of RME~\cite{Dyson1962}. He showed that the eigenvalues can be identified with the positions of charged particles interacting via two-dimensional Coulombic repulsion and executing Brownian motion. In addition a position dependent potential may be present. Dyson's log-gas model~\cite{Mehta2004,Forrester2010,Dyson1962} is of fundamental importance to the field of RMT and has been applied in various contexts~\cite{Pandey1995,Guhr1996,HJ2008,SKP1,SKP2}.  Under this formalism the three invariant ensembles of random matrices, viz. orthogonal, unitary and symplectic~\cite{Mehta2004}, turn out to be the equilibrium states of this system at three special temperatures $\beta^{-1}=1,1/2,1/4$, $k_B$ being set equal to 1. Dumitriu and Edelman, by dropping the invariance requirement, proposed matrix models which correspond to general $\beta>0$~\cite{DE2002}. Moreover, very recently it has been shown that by considering a special diffusive matrix model, invariant $\beta$-ensembles ($\beta$ restricted to certain continuous domains) may be realised~\cite{ABG2012,ABMV2012}.

In RMT one usually deals with the ensembles of unitary matrices and Hermitian matrices, which share the following generic structure for the joint probability density (JPD) of eigenvalues ($\chi_j; j=1,...,N$): 
\begin{equation}
\label{jpd_chi}
P_\chi(\{\chi\})=Z^{-1} \prod_{j<k}|\chi_j-\chi_k|^\beta \prod_{l} w(\chi_l).
\end{equation}
Here $\{\chi\}\equiv \{\chi_1,...,\chi_N\}$ and $Z=\int_{\mathbb{R}} \prod_{j<k}|\chi_j-\chi_k|^\beta \prod_{l} w(\chi_l)d\chi_l$ is the partition function, $\mathbb{R}$ being the appropriate domain of integration. $|\chi_j-\chi_k|^\beta$ is the characteristic random matrix eigenvalue-repulsion term, and $w(\chi)$ is the weight function which decides specific features of a given ensemble. Note that, for brevity, we use the term Hermitian to refer to real symmetric and self-dual quaternion matrices also. 

\begin{figure}[t]
\centerline{\includegraphics*[width=0.45\textwidth]{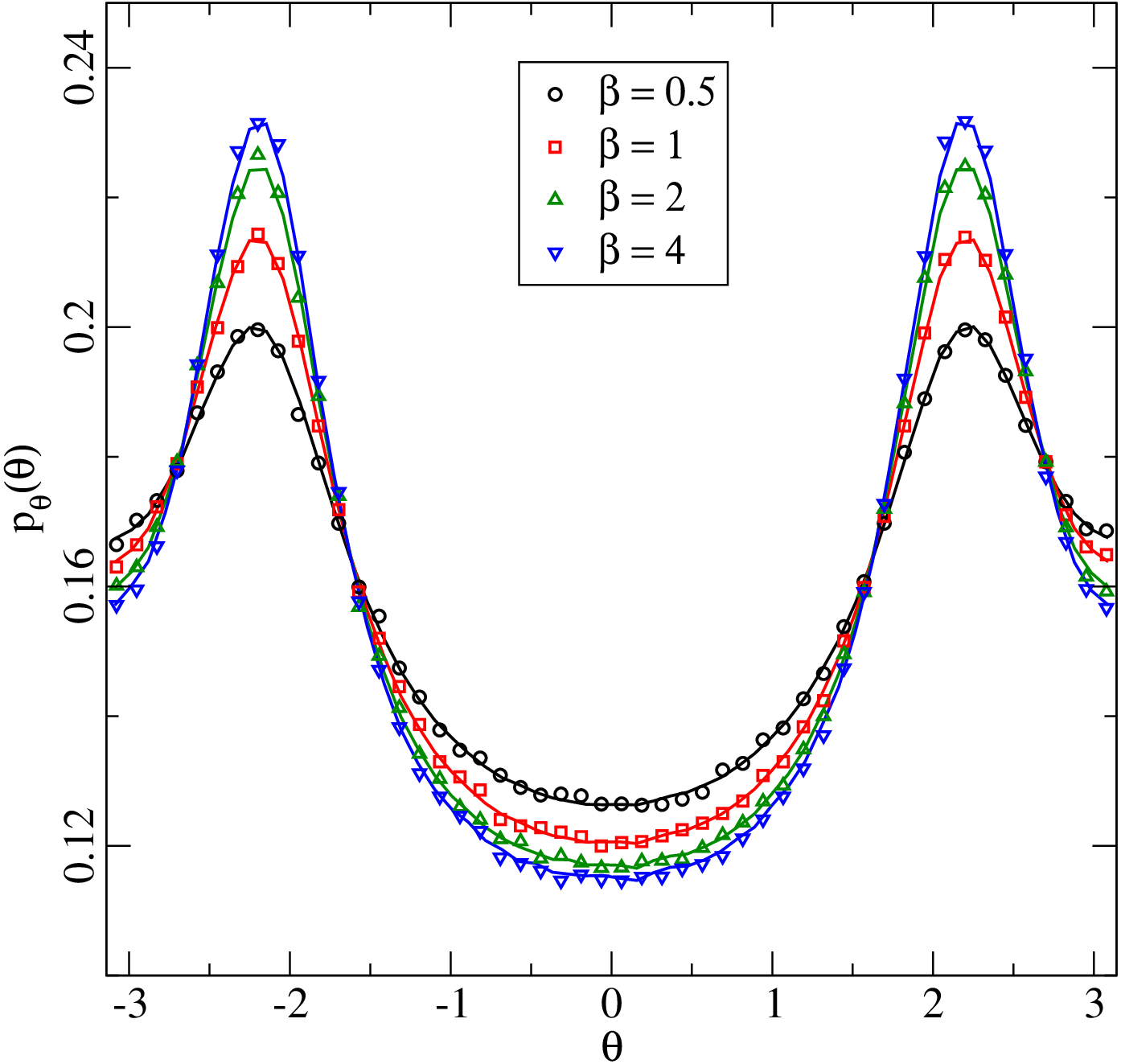}}
\caption[]{Density of eigenphases for $w(e^{i\theta})=|1 + 2e^{i\theta}+3e^{i2\theta}|^{-2\beta}$ and $N=10$. Solid lines are the result of Wand-Landau simulation. For comparison we also show the Boltzmann-sampling based Monte Carlo results (symbols). WLA gives the densities for arbitrary $\beta>0$ in a single simulation whereas the conventional Monte Carlo needs separate simulation for each $\beta$. }
\label{Fig2}
\end{figure}

In the unitary case the eigenvalues are located on a unit circle in the complex plane, i.e., they are of the form $e^{i\theta_j}$, $\theta_j\in[-\pi,\pi]$ being the eigenangles (or eigenphases). We will use $\chi$ below to refer interchangeably to $e^{i\theta}$ and $\theta$. $w(e^{i\theta})=1$ corresponds to the important case of Dyson's (uniform) circular ensemble~\cite{Mehta2004}. In the Hermitian case $\chi_j \equiv x_j$ are positions on the real line. Jacobi family of RME is obtained when $w(x)$ is chosen as weight functions associated with the classical orthogonal polynomials. For example, $w(x)=e^{-\beta x^2/2}, x^{\beta (a+1)/2}e^{-\beta x/2}$, $(1-x)^{\beta (a+1)/2} (1+x)^{\beta (b+1)/2}$ lead to the Gaussian, Laguerre and Jacobi ensembles of random matrices~\cite{Mehta2004,Forrester2010,DE2002,SKP1}.  In these three cases the  $x_j$ lie in $\mathbb{R}=(-\infty,\infty), [0,\infty)$ and $[-1,1]$ respectively. The restrictions on $a,b$  are determined from the normalisability of the weight functions.

The JPD of eigenvalues can be used to extract important information about the ensemble. For example, we can calculate the (marginal) density of eigenvalues, viz., $p_\chi(\chi)=N^{-1}\langle\sum_j\delta(\chi-\chi_j)\rangle_\mathbb{R}$. Here $\langle~\rangle_\mathbb{R}$ represents the ensemble average with respect to the JPD in eq.~\eqref{jpd_chi}. $p_\chi(\chi)$ provides the information about average behaviour of the eigenvalues and therefore reveals the general features of a given ensemble. Similarly, the distribution of an observable $A=\mathcal{A}(\{\chi\})$ depending on $\chi$'s can be obtained using $f_A(A)=\langle \delta(A-\mathcal{A}(\{\chi\})\rangle_\mathbb{R}$. 

The JPD of eigenvalues in eq. \eqref{jpd_chi} can be recast as a canonical Boltzmann-Gibbs probability density:
\begin{equation}
P_\chi(\{\chi\})=Z^{-1} e^{-\beta E(\{\chi\})}.
\end{equation}
Here $E$ represents potential energy of the system,
\begin{equation}
\label{energy}
E=-\sum_{ j<k} \ln |\chi_j-\chi_k|+\sum_{j} V(\chi_j).
\end{equation} 
In the log-gas picture the first term in the above equation represents the two-dimensional Coulombic repulsion between the charges, while $V(\chi)$ represents the local binding potential seen by the charges~\cite{Mehta2004,Forrester2010,Dyson1962,Pandey1995,SKP1}. $V(\chi)$ is related to the weight function $w(\chi)$ in eq.~\eqref{jpd_chi} as
\begin{equation}
\label{potential}
V(\chi)=-\beta^{-1}\ln w(\chi).
\end{equation}
Eq.~\eqref{energy} provides the energy function which we use in WLA to perform the random walk in $E$-space and to derive the density of states $g(E)$. $g(E)$ can eventually be used to generate the density of eigenvalues and distributions of related observables. We give the details in the next section.

It can be shown that for the above mentioned {\it classical} weight functions, $E$ is bounded from below, i.e., there exists an $E_0$ such that $E\geq E_0$ for all $\chi_j\in\mathbb{R}$ \cite{Mehta2004}. For the Dyson's circular ensemble, minimum energy configuration corresponds to the case when the angular separation between any two adjacent charges on the unit circle is equal. In this case we have \cite{Mehta2004}
\begin{equation}
E_0=-\frac{1}{2}N\ln N.
\end{equation}

For the Hermitian case (eigenvalues as positions on real line) the minimum energy configuration $\{x_1^{(0)},...,x_N^{(0)}\}$ coincides with the zeros of classical orthogonal polynomials. For the weight functions given above, $x^{(0)}$'s are precisely the zeros of Hermite: $H_N(x)$, associated Laguerre: $L_N^{(a)}(x)$, and Jacobi: $P_N^{(a,b)}(x)$ polynomials~\cite{Mehta2004,SK}. The corresponding expressions for $E_0$ are~\cite{Mehta2004,SK}
\begin{equation}
E_0=\frac{N(N-1)}{4}(1+\ln2)-\frac{1}{2}\sum_{j} j\ln j
\end{equation}
for the Gaussian case,
\begin{eqnarray}
\label{e0_lag}
\nonumber
E_0\!\!\!&=&\!\!\!\frac{N(N+a)}{2}-(N-1)\ln \Gamma(N+1)-\sum_j \frac{j-2N+2}{2}\ln j\\
&&~~~~~~~~~~~~~ -\sum_j\frac{j-1}{2}\ln(j+a)-\frac{a+1}{2}\ln \frac{\Gamma(N+a+1)}{\Gamma(a+1)}
\end{eqnarray}
for the Laguerre case, and
\begin{eqnarray}
\nonumber
E_0\!\!\!&=&\!\!\!-\frac{N(N+a+b+1)}{2}\ln2-(N-1)\ln \frac{\Gamma(N+1)\Gamma(N+a+b+1)}{\Gamma(2N+a+b+1)}\\
\nonumber
&&\!\!\!\!-\sum_j \frac{N-j}{2}\ln (N+j+a+b)-\sum_j \frac{j-2N+2}{2}\ln j-\sum_j\frac{j-1}{2}\ln[(j+a)(j+b)]\\
\nonumber
&&\!\!\!\!-\frac{a+1}{2}\ln \frac{\Gamma(N+a+1)\Gamma(N+a+b+1)}{\Gamma(a+1)\Gamma(2N+a+b+1)}
-\frac{b+1}{2}\ln \frac{\Gamma(N+b+1)\Gamma(N+a+b+1)}{\Gamma(b+1)\Gamma(2N+a+b+1)}~~~~\\
\end{eqnarray}
for the Jacobi case.

\begin{figure*}[t]
\centerline{\includegraphics*[width=0.95\textwidth]{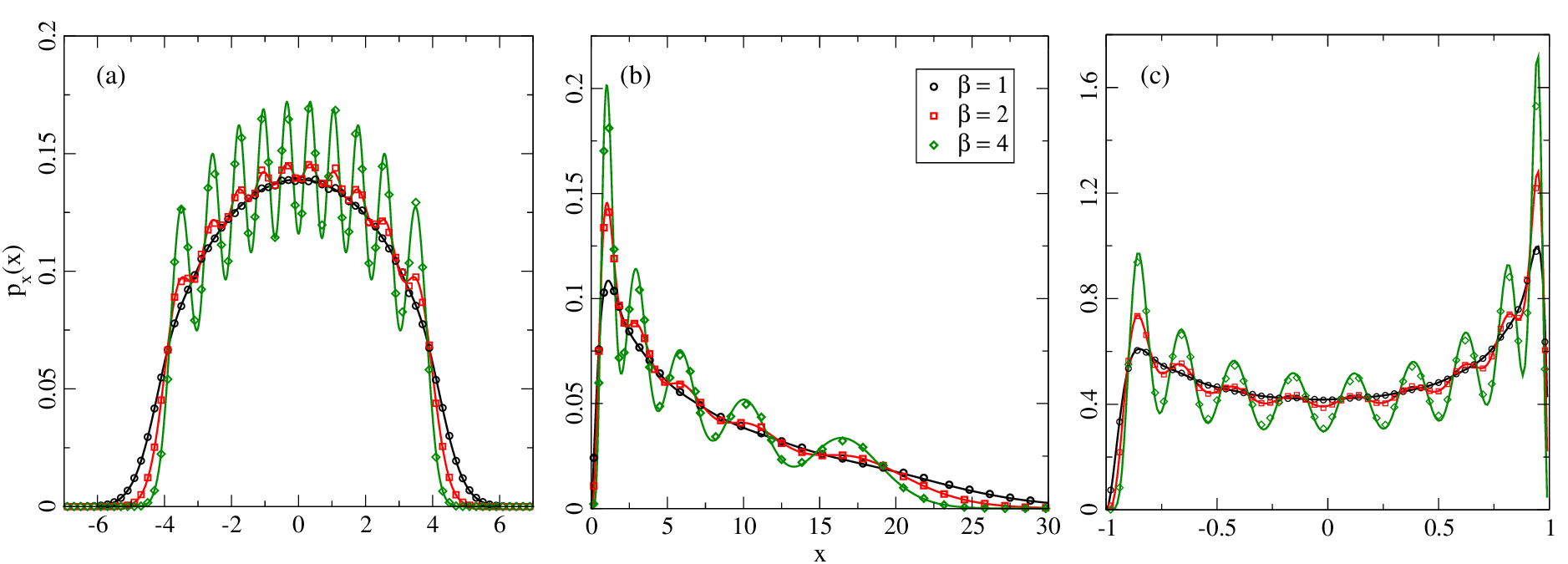}}
\caption[]{Density of eigenvalues for (a) Gaussian: $N=10$, (b) Laguerre: $N=5, a=2$, and (c) Jacobi: $N=9, a=1, b=3$ ensembles (Symbols: Simulation; Solid lines: Analytical~\cite{SKP1} ). The legend applies to all three boxes.
} 
\label{Fig3}
\end{figure*}

In the above examples for the classical ensembles we have chosen $w(\chi)$ in a way that $V(\chi)$ in eq.~\eqref{potential} is $\beta$-independent. However, often this is not the case~\cite{DE2002,Forrester2010,SKP1,SKP2}. For instance, the density of transmission eigenvalues, which govern the statistics of conductance (or transmission) in chaotic cavities, is given by~\cite{Forrester2010,SKP2}
\begin{equation}
\label{jpd_T}
P_T(\{T\})\propto \prod_{j<k}|T_j-T_k|^\beta \prod_{l}T_l^{\beta(M-N+1)/2-1}.
\end{equation}
Here $0\leq T_j\leq1$, and $M,N\,(M\geq N)$ represent the number of modes in the two ideal leads connected to the cavity. $V(T)$ in this case is $-((M-N+1)/2-1/\beta)\ln T$, which makes the energy $E$ in eq.~\eqref{energy} $\beta$-dependent too. As a consequence, for each $\beta$ we will need to do a separate simulation. This completely nullifies the usefulness of WLA over the conventional Monte Carlo scheme. However, we can circumvent this problem by considering the transformation $T_j=e^{-t_j}, 0\leq t_j<\infty$. The factor coming from the Jacobian of transformation gets rid of the unwanted $-1$ in the exponent and we obtain 
\begin{equation}
P_t(\{t\})\propto \prod_{j<k} |e^{-t_j}-e^{-t_k}|^\beta \prod_{l} e^{-\beta(M-N+1)t_l/2}.
\end{equation}
WLA can now be implemented using the $\beta$-independent energy function, eq.~\eqref{energy} with $\chi_j\equiv e^{-t_j}$ and $V=(b+1)t_j/2, b=M-N$.
Moreover, in this case
\begin{eqnarray}
\nonumber
E_0\!\!\!&=&\!\!\!-(N-1)\ln \frac{\Gamma(N+1)\Gamma(N+b)}{\Gamma(2N+b)}-\sum_j \frac{j-2N+2}{2}\ln j\\
\nonumber
&&-\sum_j \frac{N-j}{2}\ln (N+j+b-1)-\sum_j\frac{j-1}{2}\ln(j-1)\\
&&-\sum_j\frac{j-1}{2}\ln(j+b)-\frac{b+1}{2}\ln \frac{\Gamma(N+b+1)\Gamma(N+b)}{\Gamma(b+1)\Gamma(2N+b)}.
\end{eqnarray}
The term $(j-1)\ln(j-1)$ in the sum above should be taken (in a limiting sense) as 0 for $j=1$.
With the $\beta$-independent energy function we can obtain the density $p_t(t)$ for multiple values of $\beta$ in a single simulation. The density of $T$'s can then be trivially obtained using the relation $p_T(T)=e^t p_t(t)$.

As our next example we consider the JPD for $\beta$-Wishart ensemble as derived in \cite{ABMV2012}:
\begin{equation}
\label{beta_wishart}
P_\lambda(\{\lambda\})\propto \prod_{j<k} |\lambda_j-\lambda_k|^\beta \prod_{l} e^{-\frac{\lambda_l}{2}} \lambda_l^{\frac{\beta}{2}(M-N+1-\delta)-(1-\frac{\delta}{2})}. 
\end{equation}
Here $0\leq\lambda_j<\infty, \delta>0$ and $N/M\leq1$. It has been shown that for large $M,N$, a crossover between Mar\v{c}enko-Pastur and the Gamma distribution $(\beta=0)$ is observed for $\beta\sim 1/M$ \cite{ABMV2012}.  Similar to the previous example, the above JPD as such is not suitable for the implementation of WLA. We therefore carry out the transformation $\lambda_j=\beta y_j^{2/\delta}, 0\leq y_j<\infty$. This gives us the following equation for JPD of $y$'s:
\begin{equation}
P_y(\{y\})\propto \prod_{j<k} | y_j^{\frac{2}{\delta}}-y_k^{\frac{2}{\delta}} |^\beta \prod_{l}e^{-\frac{1}{2}{\beta y_l^{\frac{2}{\delta}}}} y_l^{\frac{\beta}{\delta}(M-N+1-\delta)}.
\end{equation}
This is essentially the Laguerre JPD in variables $y^{2/\delta}$ with $a$ in the effective weight function $(y^{2/\delta})^{\beta(a+1)/2} \exp(-\beta y^{2/\delta}/2) $ identified as $M-N-\delta$.
The corresponding $E$-function is given by eq.~\eqref{energy} with $\chi\equiv y_j^{2/\delta}$ and $V=2^{-1} y_j^{2/\delta}-\delta^{-1}(a+1)\ln y_j$.
The ground energy $E_0$ follows from eq.~\eqref{e0_lag} by setting $a=M-N-\delta$. The expression of density in original variables can be obtained as $p_\lambda(\lambda)=(2\beta)^{-1}\delta\, y^{1-\frac{2}{\delta}} p_y(y)$.
We would like to remark that one can also consider the case $\delta=0$ in eq.~\eqref{beta_wishart},  for which the JPD corresponds to the classical Wishart ensemble. In this case we carry out the transformation $\lambda_j=\beta e^{-y_j}, -\infty<y_j<\infty$, similar to the one considered in transmission eigenvalue JPD above.

\section{Implementation of the algorithm}

While investigation of the thermodynamic behaviour of these ensembles is interesting in itself~\cite{Mehta2004,Dyson1962,SK}, our focus here is on the evaluation of density of eigenvalues.

A successful implementation of WLA using the energy function, eq.~\eqref{energy} relies primarily on a reasonable choice of the $(E,\chi)$ space to be explored, i.e., the range (windows) for eigenvalues/eigenangles and energy. For the energy window ($W_E$), it is advantageous to shift $E$ in eq.~\eqref{energy} as $E\rightarrow E-E_0$ if $E_0$ is a priori known. Thus we perform the simulation with $E-E_0$ as the energy function with the lower end of $W_E$ set to zero. The energy being unbounded from above, fixing a reasonable upper cut off for the $W_E$ requires some hit and trial. We can get some idea from a pre Monte Carlo run where energies are generated for random configurations. This is also helpful when there is no a-priori information about $E_0$, and one has to set a reasonable lower cut off also. As an example of such a case we have considered the weight function $w(e^{i\theta})=|1+2e^{i\theta}+3e^{i2\theta}|^{-2\beta}$ in eq.~\eqref{jpd_chi}. Another option is to implement the self-adaptive method suggested in \cite{TD2005}.

\begin{figure*}[t]
\centerline{\includegraphics*[width=0.95\textwidth]{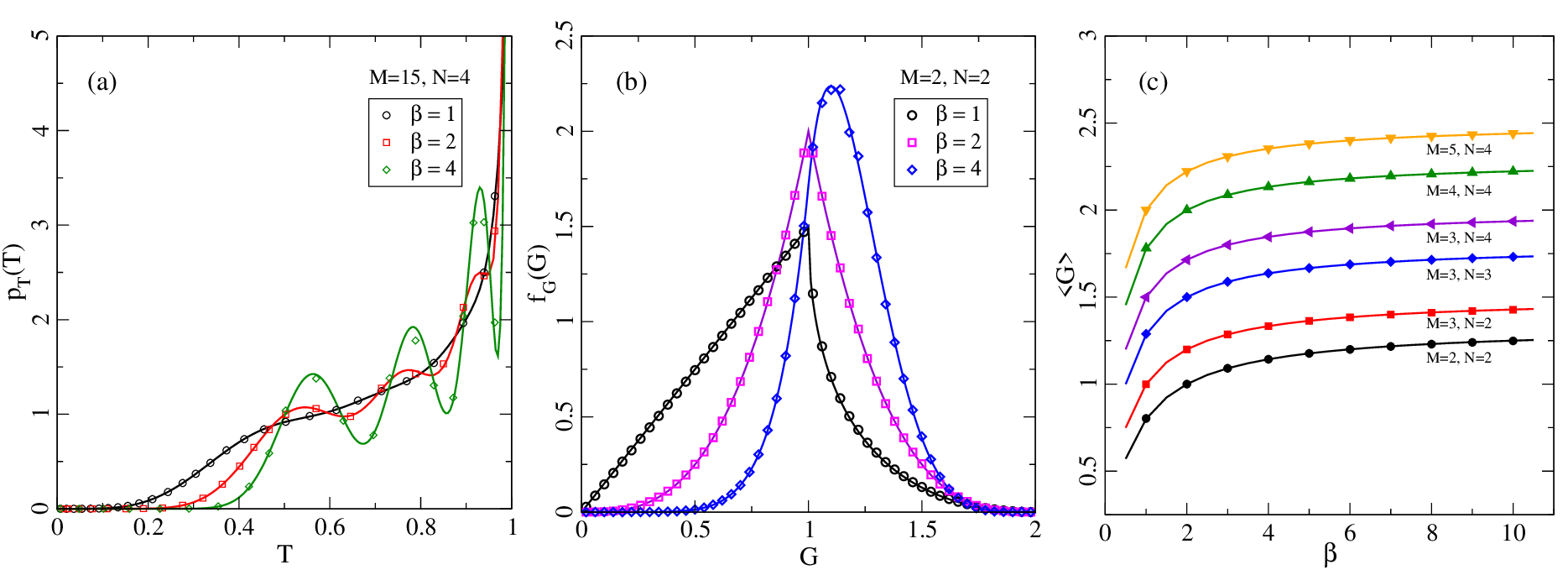}}
\caption[]{(a) Density of transmission eigenvalues for $M=15, N=4$, (b) Conductance distribution for $M=N=2$, and (c) Average conductance as a function of $\beta$ for several $M,N$ values. Symbols: Simulation; Solid lines: Analytical \cite{SKP2}.}
\label{Fig4}
\end{figure*}

The $\chi$-window $(W_\chi)$ can be trivially fixed for ensembles which have a finite support for the eigenvalues/eigenangles, e.g., Jacobi ($x_j\in[-1,1]$) and circular ($\theta_j\in[-\pi,\pi]$) ensembles. However, when the eigenvalues do not have a finite support, e.g. in Gaussian and Laguerre ensembles, we have to carefully choose the window. The choice of $W_\chi$ in these cases should be such that the $g(E)$ in the $W_E$ does not change appreciably with further increase in the $W_\chi$ width. One can also get a crude estimate using the large $N$-asymptotic result, if available. For example, in the case of Gaussian weight $e^{-\beta x^2/2}$ in eq.~\eqref{jpd_chi}, the eigenvalues give rise to Wigner semicircle of radius $\sqrt{2N}$ in the large $N$ limit. Therefore we can restrict the $W_\chi$ size to be of this order. Afterwards, the windows are divided into reasonable number of bins for sampling in the $(E,\chi)$ space. In general, too many bins will hinder the convergence, while too little bins will lead to poor results.

As mentioned before, we implement the $t^{-1}$ variant of WLA for the simulation~\cite{BP2007,BMP2008}. Once the $g(E)$ has converged to the desired accuracy~\cite{WL}, we perform the production run to obtain the joint histogram $\mathcal{H}(E,\chi)$. Owing to the symmetry of $\chi$'s we update $\mathcal{H}(E,\chi)$ by simultaneously tracking all the eigenvalues instead of just one. This speeds up building of the histogram. The density of $\chi$'s can finally be obtained using eqs.~\eqref{PA} and~\eqref{GEA}. We can also obtain the distributions or averages of multiple observables which depend on the eigenvalues by maintaining joint histogram for each of them; as has been done in some of the examples considered.
\begin{figure}[t]
\centerline{\includegraphics*[width=.45\textwidth]{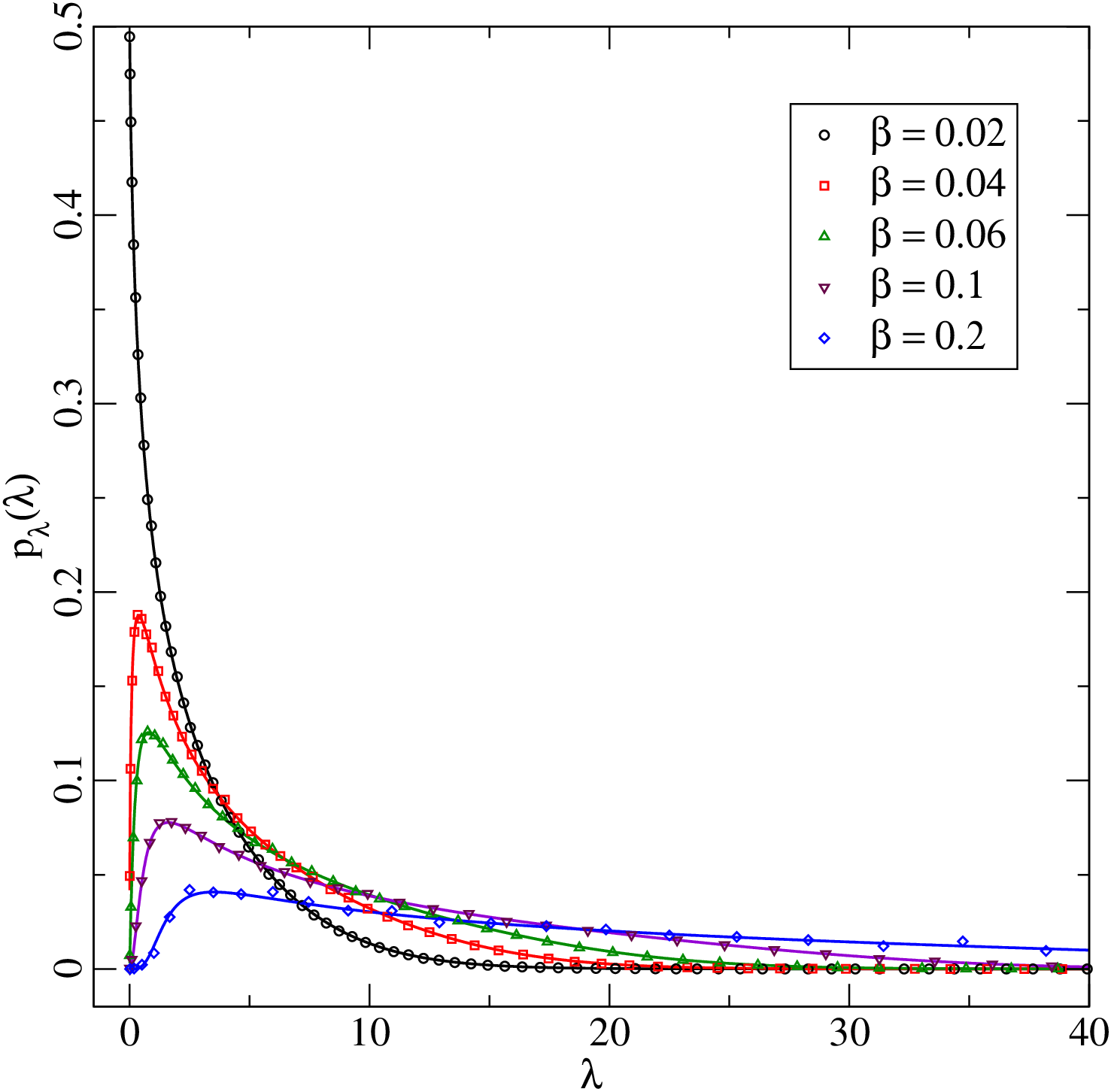}}
\caption[]{Density of eigenvalues for $\beta$-Wishart ensemble, eq.~\eqref{beta_wishart}, for $M=100, N=50$ and $\delta=1$ (Symbols: Simulation, Solid lines: Analytical~\cite{ABMV2012}). A crossover from Gamma distribution to Mar\v{c}enko-Pastur density is observed for $\beta\sim1/M$.}
\label{Fig5}
\end{figure}

\section{Results and Discussion}
\label{SecRD}

The results for Dyson's circular ensemble, $w(e^{i\chi})=1$, are shown for  $N=2$ in fig~\ref{Fig1}. Figures 1(a)-(c) show the logarithm of density of states $g(E)$, density of eigenphases $\theta_j$, and distribution of observable $A=N^{-1}\sum_j \theta_j$. The eigenangle density for each $\beta$ value is a uniform distribution~\cite{Mehta2004}. In fig. 1(c) along with the simulation results (symbols) we also show the $N=2$ analytical results (solid lines):
\begin{equation}
\label{fA}
f_A(A)=\begin{cases}
\vspace{0.2cm}

\displaystyle\frac{1}{\pi}\cos^2\displaystyle\frac{A}{2}, & \beta=1,\\

\vspace{0.2cm}

\displaystyle\frac{1}{\pi}-\displaystyle\frac{1}{2\pi^2}|2A-\sin 2A|, & \beta=2,\\

\vspace{0.2cm}

\displaystyle\frac{1}{\pi}(2-\cos A)\cos^4 \displaystyle\frac{A}{2}, & \beta=3,\\
\displaystyle\frac{1}{\pi}-\displaystyle\frac{1}{\pi^2}\bigg|A-\frac{2\sin 2A}{3}+\frac{\sin4A}{12}\bigg|, & \beta=4.
\end{cases}
\end{equation}
In fig.~\ref{Fig2}, the density of eigenphases is shown for the weight function $w(e^{i\theta})=|1 + 2e^{i\theta}+3e^{i2\theta}|^{-2\beta}$  for $N=10$. The solid lines represent the results from the Wang-Landau simulation. For comparison we also show the results of Monte Carlo simulation based on Boltzmann sampling~\cite{GPPS2003}.

Fig.~\ref{Fig3}  shows the eigenvalue densities for (a) Gaussian, (b) Laguerre and (c) Jacobi ensembles. The values of parameters considered are indicated in the caption. In fig.~\ref{Fig4}, we consider the density of transmission eigenvalues governed by JPD in eq.~\eqref{jpd_T}. Figures 4 (a), (b), (c) respectively show the density of transmission eigenvalues $T_j$, distribution of the dimensionless Landauer conductance $G=\sum_j T_j$, and average conductance $\langle G \rangle$ for several $M,N$ values. The analytical results for comparison in figs.~\ref{Fig3},~\ref{Fig4} are taken from~\cite{SKP1,SKP2}. Finally, in fig.~\ref{Fig5} we show the density of eigenvalues for the $\beta$-Wishart ensemble defined by eq.~\eqref{beta_wishart} for $M=2N=100, \delta=1$. The simulation results have been compared with the large $N$ result of~\cite{ABMV2012}. 

The time for these simulations typically varied from 5 minutes for $N\sim 5$ to 45 minutes for $N\sim50$ on a workstation with Intel Core2 6300 @1.86 GHz processor. The exact time, however, depends on the specific ensemble and the choices for the final convergence factor ($F$)~\cite{BP2007}, the $E$ and $\chi$ windows, number of bins ($n_E,n_\chi$), and the production run steps ($s_P$). For example, for fig. 3(a), ($x=\chi$), it took about 20 minutes for the choices $F=10^{-6}$, $W_E$=[0,500], $n_E=500$, $W_x$=[-7,7]$, n_x=70$, $s_P=5\times10^8$. For fig. 5, ($\lambda=\beta y^2=\beta \chi$), the time taken was about 30 minutes for $F=10^{-5}$, $W_E=[0,20000]$, $n_E=400$, $W_y=[0,45]$, $n_y=75$, $s_P=5\times10^7$.

\section{Conclusion}
\label{SecC}
We proposed a method based on the Wang-Landau algorithm for numerical generation of the eigenvalue densities and related observables of random matrix ensembles. This scheme produces the results for multiple $\beta$ values in a single simulation, and is therefore advantageous compared to the conventional Monte-Carlo scheme. This feature also makes it ideal for investigating the $\beta$-ensembles which have been of considerable interest in recent times. We demonstrated the utility of this method using several examples.

\acknowledgments
The author is grateful to T. Wirtz and T. Guhr for fruitful discussions.

\end{document}